\documentclass[10pt,conference]{IEEEtran}
\ifCLASSOPTIONcompsoc
  \usepackage[nocompress]{cite}
\else
  \usepackage{cite}
\fi
\ifCLASSINFOpdf
\else
\fi

\usepackage[english]{babel}
\usepackage{blindtext}
\usepackage[]{todonotes}
\usepackage{amsmath}
\usepackage{amssymb}
\usepackage{algorithm}
\usepackage{algpseudocode}
\usepackage{tabularx}
\usepackage{comment}
\usepackage{grffile}
\usepackage{hyperref}

\definecolor{DarkGreen}{rgb}{0.1,0.5,0.1}
\definecolor{DarkRed}{rgb}{0.5,0.1,0.1}
\definecolor{DarkBlue}{rgb}{0.1,0.1,0.5}
\definecolor{DarkPurple}{rgb}{0.5,0.2,0.5}
\definecolor{DarkTurquoise}{rgb}{0.1,0.5,0.5}
\definecolor{Red}{rgb}{1,0,0}

\begin{document}

\newcommand{\solution}{\textit{FlEC}}
\newcommand{\causalew}{FECPattern}
\newcommand{\causalth}{ds}

%
\title{FlEC: Enhancing QUIC with application-tailored reliability mechanisms}
%
%
%
%

\author{%
   \IEEEauthorblockN{François Michel\IEEEauthorrefmark{1},
                     Alejandro Cohen\IEEEauthorrefmark{2},
                     Derya Malak\IEEEauthorrefmark{3},
                     Quentin De Coninck\IEEEauthorrefmark{1},
                     Muriel M\'edard\IEEEauthorrefmark{4},
                     Olivier Bonaventure\IEEEauthorrefmark{1}}\\
   \IEEEauthorblockA{\IEEEauthorrefmark{1}%
                      UCLouvain, Belgium, \{francois.michel,quentin.deconinck,olivier.bonaventure\}@uclouvain.be}
    \IEEEauthorblockA{\IEEEauthorrefmark{2}%
                      Technion–Israel Institute of Technology, Haifa, Israel, alecohen@technion.ac.il}
    \IEEEauthorblockA{\IEEEauthorrefmark{3}%
                     Ressenlaer Polytechnic Institute,
                     New York, USA,
                     malakd@rpi.edu}
    \IEEEauthorblockA{\IEEEauthorrefmark{4}%
                     Massachusetts Institute of Technology,
                     Cambridge, USA,
                     medard@mit.edu}
}

\markboth{Journal of \LaTeX\ Class Files,~Vol.~14, No.~8, August~2015}%
{Shell \MakeLowercase{\textit{et al.}}: Bare Demo of IEEEtran.cls for Computer Society Journals}
%



\IEEEtitleabstractindextext{%
\begin{abstract}
Packet losses are common events in today's networks. They usually result in longer delivery times for application data since retransmissions are the \emph{de facto} technique to recover from such losses. Retransmissions is a good strategy for many applications but it may lead to poor performance with latency-sensitive applications compared to network coding. Although different types of network coding techniques have been proposed to reduce the impact of losses by transmitting redundant information, they are not widely used. Some niche applications include their own variant of Forward Erasure Correction (FEC) techniques, but there is no generic protocol that enables many applications to easily use them. We close this gap by designing, implementing and evaluating a new Flexible Erasure Correction (FlEC) framework inside the newly standardized QUIC protocol. With FlEC, an application can easily select the reliability mechanism that meets its requirements, from pure retransmissions to various forms of FEC. We consider three different use cases: $(i)$ bulk data transfer, $(ii)$ file transfers with restricted buffers and $(iii)$ delay-constrained messages.
We demonstrate that modern transport protocols such as QUIC may benefit from application knowledge by leveraging this knowledge in FlEC to provide better loss recovery and stream scheduling.
Our evaluation over a wide range of scenarios shows that the FlEC framework outperforms the standard QUIC reliability mechanisms from a latency viewpoint.
\end{abstract}

}

\maketitle

\IEEEdisplaynontitleabstractindextext

%
\IEEEpeerreviewmaketitle


\section{Introduction}

\IEEEPARstart{T}{he} transport layer is one of the key layers of the protocol stack. It ensures the end-to-end delivery of application data through the unreliable network layer. There are two main families of transport protocols: the unreliable datagram protocols like UDP, DCCP \cite{kohler2006designing}, RTP \cite{rfc1889} or QUIC datagrams~\cite{ietf-quic-datagram-10}, and the reliable ones such as TCP~\cite{rfc793}, SCTP~\cite{rfc4960} or QUIC \cite{langley2017quic}. 
During the last years, QUIC has attracted a growing interest thanks to its design. The QUIC specification was finalised in May 2021 \cite{rfc9000} and as of November 2021, it is already supported by more than 30M domains on the Internet~\cite{zirngibl2021s}. QUIC structures its control information and data into frames and supports stream multiplexing. Finally, QUIC includes the authentication and encryption functions of Transport Layer Security (TLS) \cite{rfc8446}. By using the latter to encrypt and authenticate all data and most of the headers, QUIC prevents interference from middleboxes. Coupled with the availability of more than two dozens open-source implementations \cite{quicwg-implems-list}, QUIC has become a very interesting platform for transport layer research~\cite{kakhki2017taking,de2019pluginizing,polese2019survey,rquic:2019,ietf-quic-recovery}. While QUIC leverages the loss recovery and congestion control techniques that are part of modern TCP implementations, other loss recovery mechanisms using FEC have recently been considered to recover earlier from packet losses~\cite{coding-for-quic,michel2019quic,rquic:2021}.

Packet losses, either caused by congestion or transmission errors are frequent in today's networks and seriously considered for the design of modern transport protocols~\cite{langley2017quic}. The first transport protocols relied on simple Automatic Repeat reQuest (ARQ) mechanisms \cite{rfc793,bertsekas1992data} to recover from losses. Over the years, a range of heuristics have been proposed. For TCP, this includes the fast retransmit heuristic~\cite{rfc6582}, selective acknowledgements~\cite{rfc2018}, the Eifel algorithm \cite{ludwig2000eifel}, recent acknowledgments~\cite{ale8985}, tail loss recovery techniques \cite{rajiullah2015evaluation,rfc5827}, and others. Other reliable transport protocols have also benefited from this effort. SCTP and QUIC include many of the optimizations added to TCP over the years \cite{budzisz2012taxonomy,langley2017quic}. 


While retransmissions remain the prevalent technique to recover from packet losses, coding techniques have been proposed in specific scenarios such as ATM networks \cite{biersack1992performance}, audio/video traffic \cite{carle1997survey} or multicast services \cite{gemmell2003pgm,rfc3452} where the cost of retransmissions grows with the group size. Some of these approaches are supported by RTP extensions~\cite{rfc2733rtpfec,rfc6682rtpfecraptor}. Several of these approaches have been applied to TCP~\cite{sundararajan2009tcpnc,kim2012ctcp}, usually by using a coding sublayer below TCP and hiding the coding functions from the transport layer \cite{medard_xors:2020,sundararajan2011network,cui2014fmtcp}.

Despite these efforts, many Internet applications still select either an unreliable transport (such as UDP) or TCP that forces in-order delivery and suffers from head-of-line blocking~\cite{langley2017quic}. If an application developer needs another reliability model, she needs to implement the logic directly inside the application.

In this article we propose to revisit the reliability mechanisms in the transport layer. Our main contribution is that we enable applications to finely tune the reliability mechanism of the transport protocol to closely fit their needs. We implement our solution using QUIC and protocol plugins~\cite{de2019pluginizing}, but our ideas are generic and can be applied to other protocols as well. We evaluate the flexibility of our techniques by considering a range of applications and show that our application-tailored reliability mechanism outperforms a one-size-fits-all solution.

This paper is organised as follows. We first discuss the current reliability mechanisms (Section~\ref{section:reliability}) of transport protocols and see how flexibility is currently provided by existing solutions (Section~\ref{section:tunable}). We then propose Flexible Erasure Correction (\solution{}), a novel reliability mechanism that can be easily redefined on a per-application basis (Section~\ref{section:flec}) to adapt the reliability mechanism to the application needs. We implement \solution{} inside QUIC (Section~\ref{section:implementation}) and demonstrate the benefits of the approach by studying three different use-cases (Sections~\ref{section:bulk}-\ref{section:messaging}) with competing needs that can all improve their quality of experience using \solution{}.

\section{Related work}\label{section:reliability}

The QUIC protocol currently provides a reliable bytestream abstraction using \textit{streams}. 
Application data transiting through QUIC streams is carried inside \texttt{STREAM} frames. Upon detection of a loss, the application data carried by the lost \texttt{STREAM} frames are retransmitted in new \texttt{STREAM} frames sent in new QUIC packets. 
QUIC~\cite{ietf-quic-recovery} uses two thresholds to detect losses: a \textit{packet-based} threshold and a \textit{time-based} threshold. The packet-based threshold marks a packet as lost after packets with a sufficiently higher packet number have been received, indicating that pure re-ordering is unlikely. The specification recommends that an unacknowledged packet with number $x$ is marked as lost when the acknowledgement of a packet with a number larger or equal to $x+3$ has been received. This is similar to TCP's fast retransmit heuristic. The time-based threshold marks an unacknowledged packet as lost after a sufficient amount of time when a packet sent later gets acknowledged. 
The specification recommends that an unacknowledged packet sent at time $t$ is marked as lost after $t + \frac{9}{8}*RTT$ if a packet sent later has already been acked. These two thresholds are only reached after at least one round-trip time, resulting in a late retransmission for delay-constrained applications. Such applications would benefit from a reliability mechanism that corrects packet losses \textit{a-priori}.

There are ongoing discussions within the IETF \cite{coding-for-quic} and the research community \cite{michel2019quic,de2019pluginizing,rquic:2019} to add Forward Erasure Correction (FEC) capabilities to QUIC. 
This mechanism consists in sending redundant information (Repair Symbols) before packets (Source Symbols) are detected as lost.
It is especially useful for applications that cannot afford to wait for a retransmission, either due to strong delay requirements or connections suffering from long delays. Google experimented with a naive XOR-based FEC solution in early versions of QUIC \cite{swett-quic-fec}. The IRTF Network Coding research group explored alternative solutions~\cite{coding-for-quic}. Relying on a XOR code~\cite{swett-quic-fec} does not enable sending several repair symbols to protect a window of packets, preventing the solution from recovering loss bursts.  The standardisation work~\cite{coding-for-quic} does not provide any performance evaluation nor technique to schedule source and repair symbols.  QUIC-FEC~\cite{michel2019quic} proposes several redundancy frameworks and codes for the QUIC protocol. In this previous work, we studied file transfers with different codes such as XOR, Reed-Solomon and Random Linear Codes (RLC). We showed that FEC with QUIC can be beneficial for small file transfers but is harmful for longer bulk transfers compared to Selective-Repeat ARQ (SR-ARQ) mechanisms. One of the main limitations of QUIC-FEC~\cite{michel2019quic} is that the code rate is fixed during the connection, leading to the sending of unnecessary coded packets. 
Pluginized QUIC (PQUIC)~\cite{de2019pluginizing} proposes a FEC plugin equivalent to the RLC part of QUIC-FEC.  Finally, rQUIC~\cite{rquic:2021} presents an adaptive algorithm to regulate the code rate in function of the channel loss rate for QUIC communications. rQUIC has two main differences with our work. First, rQUIC assumes that isolated losses are not due to congestion. When it recovers isolated lost packets, rQUIC hides their loss signal to QUIC's congestion control to benefit from a larger bitrate. We follow the IRTF NWCRG recommendations~\cite{irtf-nwcrg-coding-and-congestion-09}: we never hide any signal to the congestion control. If isolated losses are not due to congestion, then a specific congestion control ignoring isolated losses can be used instead of hiding the loss signal. The second difference is that we adapt the redundancy both to the channel conditions and the application requirements. Our solution will neither send the same amount of redundancy nor the same pattern of Repair Symbols for a bulk download and for a real-time video conferencing application. 
%

Network coding has been considered for other transport protocols \cite{rfc5109,ietf-rtcweb-fec,cavusoglu2003real}. 
TCP/NC~\cite{sundararajan2009tcpnc} adds network coding to TCP connections by applying a coding layer beneath the transport layer. It improves the TCP throughput by recovering from packets losses that block the TCP window. CTCP~\cite{kim2012ctcp} pushes the idea further and proposes a revised congestion control algorithm for wireless communications. Tetrys~\cite{tournoux2011fly} proposes a coding mechanism focused on real-time video applications and develops heuristics to adjust the coding rate to the sender's behaviour. RFC5109~\cite{rfc5109} defines a standard RTP packet format to allow the use of FEC for RTP applications.
An IETF draft~\cite{ietf-rtcweb-fec} presents guidelines and requirements for the use of FEC for protecting video and audio streams in WebRTC. Minion~\cite{nowlan2012fitting} also uses coding to support unreliable data transfer above TCP.
Existing work propose multi-path solutions to handle links with poor delay and fluctuating bandwidth~\cite{kuhn2014daps,chiariotti2021hop,garcia2017low} and use FEC to reduce head-of-line blocking.
Unfortunately, none of the current solutions adapts the reliability mechanism to different classes of applications.

\section{Tunable reliability mechanisms}\label{section:tunable}

Sending Repair Symbols for delay-sensitive applications is done at the cost of bandwidth when there is no loss to recover.  This is why Repair Symbols should be sent carefully to avoid consuming bandwidth with no or low additional benefit.


Some adaptive FEC mechanisms have been proposed to adjust the redundancy overhead to the measured loss rate. Both CTCP~\cite{kim2012ctcp} and TCP/NC~\cite{sundararajan2009tcpnc} adjust the level of redundancy according to the measured loss rate. rQUIC~\cite{rquic:2021} proposes a similar idea. While these approaches can show significant benefits compared to classical retransmission mechanisms, they still increase the overhead compared to the more efficient selective-repeat mechanisms in bulk download use-cases. By using a causal scheduling algorithm, we allow our solution to react to the current channel condition and adopt similar behaviours to SR-ARQ when it is needed by the use-case.
On the other-side, real-time applications cannot fully leverage the benefits of such transport-layer coding mechanisms because there is no way for an application to precisely express its requirements. The result is that such applications typically implement their own coding-enabled protocol~\cite{rfc3550rtp,rfc2733rtpfec}.

We reconcile strong application delay requirements and regular transport protocols by providing them with a tunable reliability mechanism that applications can adapt to their needs with small to no effort. 
We use QUIC to demonstrate our ideas, but they could be applicable to other transport protocols as well. 
QUIC stacks are mostly implemented as libraries that can be used by a wide range of applications.
While QUIC can easily be tuned on the server-side to better fit the application requirements, obtaining such a flexibility on the end-user devices is more complicated, as the application wants to tune the underlying stack to meet its requirements.
In the TCP/IP stack, this tuning is mainly done by using socket options or system-wide parameters. Socket Intents~\cite{schmidt2013socketintents} and the ongoing work~\cite{ietf-taps-arch-06} within the IETF TAPS working group show that there is an interest to insert some knowledge from the application to the transport protocol.
The QUIC specification~\cite{ietf-quic-transport} does not currently define a specific API between the application and the transport protocol but specifies a set of actions that could be performed by the application on the streams (e.g. reading and writing data on streams) and on the connection itself (e.g. switching on/off 0-RTT connection establishment or terminating the connection). The QUIC specification allows the application to pass information about the relative priority of the streams. However, it is unclear how, e.g., an application could express timeliness constraints. 

On the other side, the current FEC specification for QUIC~\cite{coding-for-quic} does not 
guide the application to choose a code rate nor which parts of the application data should be FEC-protected and when coded symbols should be sent. Furthermore, different applications may require different strategies to send redundancy.
In a video-conferencing application, Repair Symbols could protect a whole video frame. An IoT application~\cite{eggert2020quiciot} with limited buffers may want to protect the data incrementally to ensure a fast in-order delivery despite losses.

\paragraph*{Contributions}
Previously~\cite{cohen2020adaptive}, we provided a joint coding scheme and algorithm in which one can theoretically manage the delay-rate tradeoff to get the required QoS. However, 
we did not cope with the complex and various requirements of real applications.
We advocate that sending redundancy packets in transport protocols \textbf{should be done in adequacy with the application needs in order to provide satisfactory results.}
In previous solutions, the FEC mechanism does not track both the channel condition and application requirements throughout the data transfer. In this work, we consider both the application's requirements and the network conditions to schedule the redundancy more efficiently. 
The contribution of this article is thus a complete redefinition of the reliability mechanism of the QUIC protocol by making it general and flexible. We introduce a general loss recovery framework able to implement both a classical SR-ARQ mechanism and FEC.  
We leverage the idea of protocol plugins~\cite{de2019pluginizing} and implement our reliability mechanism as a framework exposing two anchors points to applications. Applications can redefine these anchors according to their delay-sensitivity and traffic pattern. We explore three different use-cases and show that adapting the reliability mechanism to the use-case can drastically improve the quality of the transmission. The first use-case is the bulk download scenario, discussed in Section~\ref{section:bulk}. The second use-case discussed in Section~\ref{section:buffer_limited} is a scenario where the peer's receive window is small, resulting in the sender being regularly blocked by the flow control during loss events. The third use-case discussed in Section~\ref{section:messaging} is a scenario where the application sends messages that must arrive before a specific deadline. The three use-cases are described below.

\subsection{Bulk file transfer}

Bulk file transfer is the simplest use-case we consider. It consists in the download of a single file under the assumption that the receive buffer is large compared to the bandwidth-delay product of the connection. This is the classical use case for many transport protocols. Current open-source QUIC implementations use default receive window sizes that support such a use-case. The receive window starts at 2 MBytes for locally-initiated streams in \texttt{picoquic}~\cite{picoquic-flowcontrol-limit}. The Chromium browser's implementation~\cite{chromium-flow-control-limit} 
starts with an initial receive window of 6MB per stream and 16MB for the whole connection. The metric that we minimize here is the total time to download the whole file. This includes REST API messages
that often need to be completely transferred in order to be processed correctly by the application.
As already pointed out~\cite{flach2013reducing,michel2019quic,langley2017quic}, a packet loss during the last round-trip-time can have a high relative impact on the download completion time. The latter may indeed be doubled for small files due to the loss of a single packet. Protecting these tail packets can drastically improve the total transfer time at a cost significantly smaller than the cost of simply duplicating all these packets.
On the other hand, protecting other packets than the tail ones
with FEC can be harmful for the download completion time. The packet losses in the middle of the download can be recovered without FEC before any quiescence period provided that the receiver uses sufficiently large receive buffers.

\subsection{Buffer-limited file transfers}
In numerous network configurations, the available memory on the end devices is a limiting factor. It is common to see delays longer than 500 milliseconds in satellite communications, while their bandwidth is in the order of several dozens of Megabits per second~\cite{thomas2019quicsat,kuhn-quic-4-sat}. Furthermore, with the arrival of 5G, some devices will have access to bandwidth up to 10Gbps~\cite{rappaport2013millimeter,agiwal2016next}.
While the edge latency of 5G infrastructures is intended to be in the order of a few milliseconds~\cite{3gpp.21.915}, the network towards the other host during an end-to-end transport connection may be significantly higher, partially due to the large buffers on the routers and the buffer-filling nature of currently deployed congestion control mechanisms.
Packet losses occurring on those high Bandwidth-Delay Product (BDP) network configurations imply a significant memory pressure on reliable transport protocols running on the end devices.
To ensure an in-order delivery, the transport protocol running on these devices needs to keep the data received out-of-order during at least one round-trip-time, requiring receive buffer sizes to grow to dozens of megabytes for each connection.
At the same time, QUIC is also considered for securing connections on IoT devices~\cite{eggert2020quiciot,kumar2019quicmqtt}. Those embedded devices cannot dedicate large buffers for their network connections.
Receive buffers that cannot bear the bandwidth-delay product of the network they are attached to are unable to fully utilize its capacity, even without losses. This typically occurs when the receive window is smaller than the sender's congestion window. Measurements show that TCP receivers frequently suffer from such limitations \cite{langley2017quic}. The problem gets even worse in case of packet losses as they prevent the receiver to deliver the data received out-of-order to the application. Those data will remain in memory, reducing the amount of new data that can be sent until the lost data is correctly retransmitted and delivered to the application.
Sacrificing a few bytes of the receiver memory in order to handle repair symbols and protect the receive window from being blocked upon packet losses can drastically improve the transfer time, even in a file transfer use-case.
In such cases, FEC can be sent periodically along with non-coded packets during the download and not just at the end of the transfer.

\subsection{Delay-constrained messaging}

Finally, we consider applications with real-time constraints such as video conferencing.
Those applications send messages (e.g. video frames) that need to be successfully delivered within a short amount of time. The metric to optimize is the number of messages delivered on-time at the destination.

FEC can significantly improve the quality of such transfers by recovering from packet losses without retransmissions, at the expense of using more bandwidth. Researchers have already applied FEC to video applications~\cite{cavusoglu2003real,puri2001forward}. Some~\cite{cavusoglu2003real} take a redundancy rate as input and allocates the Repair Symbols given the importance of the video frame. Others~\cite{puri2001forward} propose a congestion control scheme that reduces the impact of isolated losses on the sending rate. They then use this congestion control to gather knowledge from the transport layer to the application in order to adapt the transmission to the current congestion. We propose the reverse idea: the application transfers its knowledge directly in the transport protocol to automatically adjust its stream scheduler and redundancy rate given the application's requirements.

\section{FlEC}\label{section:flec}

\begin{figure}
    \centering
\includegraphics[trim=0cm 0.2cm 0cm 0cm,width=0.9\linewidth]{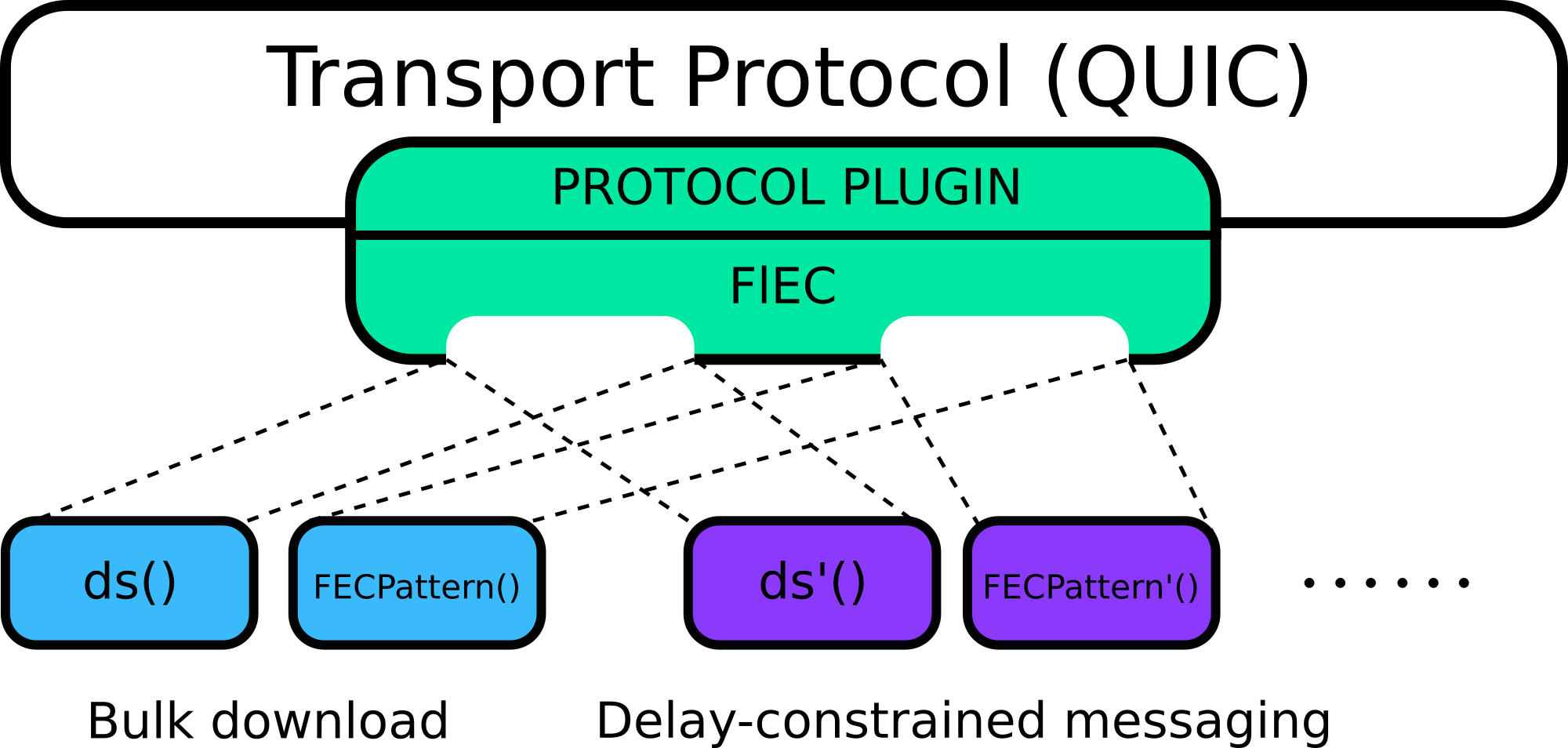}
    \caption{Design of the solution: a general framework with two pluggable anchors to redefine the reliability mechanism given the use-case.}
    \label{fig:design}
\end{figure}

In this section, we present the Flexible Erasure Correction (\solution{}) framework. \solution{} starts from a previous theoretical work, AC-RLNC~\cite{cohen2020adaptive}. This previous work proposes a decision mechanism to schedule repair symbols depending on the network conditions and the feedback received from the receiver. In this approach, repair symbols are sent in reaction to two thresholds:
the first is triggered as a function of the number of missing degrees of freedom by the receiver, and the second threshold sends repair symbols \textit{once every RTT}. The original goal of the proposed algorithm~\cite{cohen2020adaptive} is to trade bandwidth for minimizing the in-order delivery delay of data packets. 

We start from this idea of tracking the sent, seen and received degrees of freedom as a first step to propose a redundancy scheduler for the transport layer.
However, while this first idea provides a general behaviour, this may be insufficient for real applications with tight constraints that cannot be expressed with AC-RLNC's parameters. For example, a video-conferencing application may prefer to maximize bandwidth over low-delay links and therefore rely on retransmissions only, while FEC is needed over high-delay links as such retransmissions cannot meet the application's delay constraints.
Instead of proposing configurable constant thresholds to tune the algorithm, we make it dynamic by proposing two redefinable functions: $\causalth{}()$ (for ``\textit{delay-sensitivity}'') and $\causalew{}()$. These two functions can be completely redefined to instantiate a reliability mechanism closely corresponding to the use-case. This allows having completely different FEC behaviours for use-cases with distinct needs such as HTTP versus video-conferencing. The $\causalth{}()$ threshold represents the sensitivity of the application to the in-order delivery delay of the data sent. In AC-RLNC, the FEC scheduler sends redundancy once per RTT. In \solution{}, the $\causalew{}()$ dynamic function allows triggering the sending of FEC at specific moments of the transfer depending on the use-case. Sending FEC for every RTT may deteriorate the application performance, especially when the delay is low enough to rely on retransmissions only. Having a dynamic $\causalew()$ function avoids this problem. For instance, in a bulk download scenario, it can trigger FEC at the end of the download only and rely on retransmissions otherwise. For video transfer, it can trigger FEC after each video frame is sent. Figure~\ref{fig:design} illustrates the idea of \solution{}. The regular QUIC reliability mechanism is based on SR-ARQ. In \solution{}, the SR-ARQ mechanism is a particular case among many other possibilities. Algorithm~\ref{causalRLNCalgo} shows our generic framework and Table~\ref{tab:symbols} defines the variables used by our algorithms. We implement \solution{} using PQUIC~\cite{de2019pluginizing} and define $\causalew{}()$ and $\causalth{}()$ as protocol operations. However, the same principles can be applied without PQUIC with the application redefining the operations natively thanks to the user-space nature of QUIC.

\begin{table}
    \centering
    \begin{tabularx}{0.9\linewidth}{|c|X|}
    \hline
        $\hat{l}$ & the estimated uniform loss rate \\
    \hline
        $\hat{r}$ & the estimated receive rate \\
    \hline
        $\hat{G_p}$, (resp. $\hat{G_r}$) & the estimated transition probability from the GOOD to the BAD (resp. BAD to GOOD) state of a Gilbert loss model~\cite{gemodel} \\
    \hline
        $md$ & missing degrees of freedom\\
    \hline
        $ad$ & added degrees of freedom \\
    \hline
        $\causalth{}()$ & customizable threshold eliciting Repair Symbols given the application's delay sensitivity \\
    \hline
        $\causalew{}()$ & customizable condition to send FEC using the application's traffic pattern \\
    \hline

    \end{tabularx}
    \caption{Definition of the different symbols.}
    \label{tab:symbols}
\end{table}

\begin{algorithm}\small
\begin{algorithmic}[1]
\Require $\hat{l}$
\Require $feedback$, the most recent feedback received from the peer
\Require $W$, the current coding window
\State $\hat{r} \gets 1-\hat{l}$
\State $ad \gets computeAd(W)$
\State $md \gets computeMd(W)$
\If{$feedback = \varnothing$}
    \If{\causalew()}
        \State \Return $NewRepairSymbol$
    \Else
        \State \Return $NewData$
    \EndIf
\Else
    \State updateLossEstimations(feedback)
    \If{\causalew()}
        \State \Return $NewRepairSymbol$
    \ElsIf{$\hat{r}-\frac{md}{ad} < \causalth()$}
        \State \Return $NewRepairSymbol$
    \Else
         \State \Return $NewData$
    \EndIf
\EndIf
 \caption{Generic redundancy scheduler algorithm. The $\causalth{}()$ and $\causalew{}()$ thresholds are redefined by the underlying application. The algorithm is called at each new available slot in the congestion window of the protocol.\label{causalRLNCalgo}}
\end{algorithmic}
\end{algorithm}

The $computeMd$ function computes the number of missing degrees ($md$) of freedom (i.e. missing source symbols) in the current coding window. The $computeAd$ function computes the number of added degrees ($ad$) of freedom (i.e. repair symbols) that protect at least one packet in the current coding window. Compared to AC-RLNC, we only consider in-flight repair symbols in $ad$ to support retransmissions when repair symbols are lost. 
The higher the value returned by $\causalth{}()$, the more likely it is to send repair symbols prior to the detection of a lost source symbol and the more robust is the delay between the sending of the source symbols and their arrival at the receiver.
The extra cost is the bandwidth utilization.
Sending repair symbols \emph{a priori} occupies slots in the congestion window and is likely to increase the delay between the generation of data in the application and its actual transmission. Setting $\causalth{}()$ to $-\hat{l}$ triggers the transmission of repair symbols only in reaction to a newly lost source symbol, implementing thus a behaviour similar to regular QUIC retransmissions. In this work, retransmissions are done using repair symbols to illustrate that the approach is generic. However, regular uncoded retransmissions can be used for better performance without loss of generality.
$\causalew{}()$ allows regulating the transmission of \emph{a priori} repair symbols regardless of the channel state, in contrast with AC-RLNC~\cite{cohen2020adaptive} where this threshold is triggered once per RTT.

Table~\ref{tab:use_cases} describes how $\causalth{}()$ and $\causalew{}()$ can be redefined to represent reliability mechanisms that fit the studied use-cases. The first row of the table shows how to implement the classical Selective-Repeat ARQ mechanism used by default in QUIC. The second one implements the behaviour of AC-RLNC~\cite{cohen2020adaptive}. $\causalew{}$ is triggered once every RTT according to the EW parameter of AC-RLNC. The third one is tailored for the bulk use-case: $\causalth{}()$ is set to send Repair Symbols only when there are missing symbols at the receiver and $\causalew{}()$ sends Repair Symbols when there is no more data to send. The two other rows are explained in details in the next sections. In this Table, $c$ is a non-negative user-defined constant. The higher $c$ is, the more sensitive we are to a variance in the loss rate.


\begin{table}
    \centering
    \begin{tabularx}{0.9\linewidth}{|c|c|X|}
        \hline
         \textbf{Use-case} & \textbf{$\causalth{}()$} & \textbf{$\causalew{}()$} \\
         \hline
         Bulk transfer (SR-ARQ) & $-\hat{l}$ & $false$ \\
         \hline
         AC-RLNC~\cite{cohen2020adaptive} & $c\cdot\hat{l}$ & $true$ every RTT \\
         \hline
         Bulk transfer & $-\hat{l}$ & $allStreamSent()$ \\
         \hline
         Buffer-limited bulk & $c\cdot\hat{l}$& Algorithm~\ref{algo:buffer_ew} \\
         \hline
         Messaging & $-\hat{l}$ & Algorithm~\ref{algo:ew_messages} \\
         \hline
    \end{tabularx}
    \caption{Definition of $\causalth{}()$ and $\causalew{}()$ for the considered use-cases. }
    \label{tab:use_cases}
\end{table}


\subsection{Comparing \solution{} and previous work}
The origin of \solution{} comes from the shortcomings of AC-RLNC~\cite{cohen2020adaptive} and QUIC-FEC~\cite{michel2019quic}.
As said earlier in this section, \solution{} shares with AC-RLNC the idea of tracking the state of the communication in terms of received, seen and lost symbols. However, it adds the tight and diverse application requirements to the loop in order to adopt a correct behaviour for use-cases where FEC can be beneficial. It also adds all the transport-layer considerations such as staying fair to the congestion control of the protocol upon loss recovery.

\solution{} also builds upon QUIC-FEC as it integrates similar transport layer considerations. For instance, \solution{} uses as similar wire format as well as the concept of \texttt{RECOVERED} frame in order to differentiate packet acknowledgements from symbols recoveries. However, QUIC-FEC was designed without any care of the application traffic pattern or channel condition: the packet redundancy was not adaptive at all.

\section{Implementation}\label{section:implementation}

\solution{} is composed of two parts. First, the general \solution{} framework allows defining reliability mechanisms in a flexible way. This part is generic and is not intended to vary. 
The second part contains the $\causalew{}()$ and $\causalth{}()$ operations. These operations are designed to vary depending on the use-case, so the app can redefine them based on their requirements. 

We implement our \solution{} framework inside PQUIC~\cite{de2019pluginizing}. We implement the behaviours of the three use-cases discussed in this article by redefining $\causalew{}()$ and $\causalth{}()$ to support the adequate reliability mechanism for each of them.
Similarly to previous works~\cite{michel2019quic,cohen2020adaptive}, we rely on random linear codes for the encoding and decoding of the symbols. This choice is made out of implementation convenience although other error correcting codes can be used as encoding/decoding tools of our work with only little adaptation. We advocate that even simpler codes such as Reed-Solomon can provide benefits for the considered use-cases although the benefit may be lower (e.g. such simple block codes cannot mix the repair symbols of different generations conversely to random linear codes).

We re-implemented 
the FEC plugin originally proposed in PQUIC~\cite{de2019pluginizing} to match the latest design of the FEC extension for QUIC~\cite{coding-for-quic}. We enhanced the $GF(2^8)$ RLC implementation to use dedicated CPU instructions and adding an online system solver for faster symbols recovery. Most of the \solution{} protocol operations consist in monitoring the current packet loop and providing a shim layer between the PQUIC design and the \solution{} symbols scheduling algorithm. While we propose the \solution{} framework as a protocol plugin, it can also be implemented natively and provided by default with the protocol implementation. The application can also provide its native implementation for the $\causalth{}()$ and $\causalew{}()$ operations. 
The whole \solution{} framework implementation takes 8200 lines of code. It adds a complete FEC extension to the QUIC protocol with the RLC error correcting code using PQUIC protocol plugins. This code is generic and does not have to be redefined by any application.
The codes needed to define $\causalth{}()$ and $\causalew{}()$ for the bulk and buffer-limited use-case have been written with respectively 57 and 97 lines of C code while the code for the messaging use-case takes 335 lines of C code. These two small functions are the parts that can be redefined by the application to stick to their use-case. Applications can also use our implementations for the three use-cases explored in this paper.

\section{Bulk file transfers}\label{section:bulk}
We here present the implementation and evaluation of the reliability mechanism proposed for bulk file transfers. The metric to minimize is the total download completion time. Sending unneeded repair symbols reduce the goodput and increase the download completion time. The expected behaviour is therefore similar to SR-ARQ with tail loss protection. The Repair Symbols are always sent within what is allowed by the congestion window, meaning that \solution{} does not induce any additional link pressure.

\subsection{Bulk reliability mechanism}

For a file transfer, we set the delay-sensitivity threshold to be equal to $-\hat{l}$.
\begin{equation} \label{equation_bulk_l}
\textstyle \hat{r} - md/ad < -\hat{l} \rightarrow sendRepairPacket()
\end{equation}
Substituting $\hat{l}$ by $(1-\hat{r})$ in Equation~\ref{equation_bulk_l}, we can rewrite it as
\begin{equation} \label{equation_bulk_r}
\textstyle md/ad > 1 \rightarrow sendRepairPacket()
\end{equation}
so that we send Repair Symbols only when a packet is detected as lost and it has not been protected yet. The transmission of a Repair Symbol triggered by this threshold increases $ad$ by 1 until $ad$ becomes equal to $md$.
Using the threshold defined in Equation~\ref{equation_bulk_l} ensures a reliable delivery of the data but does not improve the download completion time in the case of tail losses. $md$ only increases after a packet is marked as lost by the QUIC loss detection mechanism. The $\causalew{}()$ operation controls the \emph{a priori} transmission of Repair Symbols. In contrast with the previous solution~\cite{cohen2020adaptive}, we redefine $\causalew{}()$ and set it to $true$ only when all the application data has been sent instead of setting it to $true$ once per RTT. This implies that only the last flight of packets will be protected. All the previous flights will be recovered through retransmissions. Indeed, given the fact that the receive window is large enough compared to the congestion window, there will be no silence period implied by any packet loss except for the last flight of packets. The total download completion time will thus not be impacted by any loss before the last flight of packets. Without using FEC, the loss of any packet in the last flight will cause a silence period between the sending of that lost packet and its retransmission.
We track the loss conditions throughout the download and trigger the $\causalew{}()$ threshold according to the observed loss pattern. This loss-rate-adaptive approach is especially beneficial when enough packets are exchanged to accurately estimate the loss pattern. This occurs when the file is long or when loss information is shared among connections with the same peer.  When a sufficient number of repair symbols are sent to protect the expected number of lost source symbols, the algorithm keeps slots in its congestion window to transmit new data. Another approach would be to define $\causalew{}()$ to use all the remaining space in the congestion window to send repair Symbols, with the drawback of potentially consuming more bandwidth than needed.

\subsection{Evaluation}

We now evaluate \solution{} with the $\causalth{}$() and $\causalew{}()$ protocol operations defined for the bulk use-case.
\subsubsection{Experimental setup}
\begin{figure}
    \centering
\includegraphics[trim=0cm 0.2cm 0cm 0cm,width=0.7\linewidth]{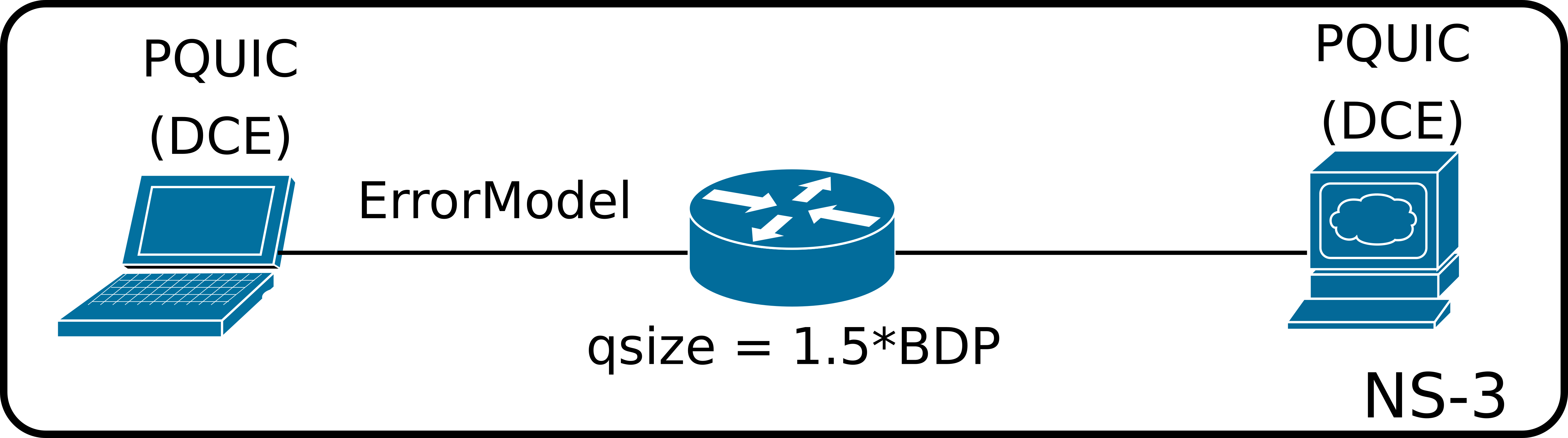}
    \caption{Experimental topology using NS-3 with Direct Code Execution.}
    \label{fig:exp_topo}
    \vspace{-0.5cm}
\end{figure}

We base our implementation on the PQUIC~\cite{de2019pluginizing} pluginized QUIC implementation on commit \textit{68e61c5}~\cite{pquic-github}. PQUIC is itself based on the \texttt{picoquic}~\cite{picoquic} QUIC implementation.
We perform numerous experiments and compare it with the regular QUIC without our plugins. We use ns-3~\cite{riley2010ns3} version 3.33 with the Direct Code Execution (DCE)~\cite{camara2014dce} module. The DCE module allows using ns-3 with the code of a real implementation in a discrete time environment. This means that the actual code of the QUIC and \solution{} implementation is running and that the underlying network used by the implementation is simulated by ns-3, making the experiments fully reproducible while running real code. Figure~\ref{fig:exp_topo} shows the experimental setup. We use ns-3's \textit{RateErrorModel} to generate reproducible loss patterns with different seeds and configure the network queues to 1.5 times the bandwidth-delay product.
We run the system into a Ubuntu 16.04 Linux system with 20GB of RAM, using 16 cores \textit{Intel(R) Xeon(R) Silver 4314} CPUs. 

Although the congestion control is orthogonal to our proposed reliability mechanism, the Reno~\cite{newreno} and CUBIC~\cite{ha2008cubic} congestion control algorithms supported by PQUIC suffer from bandwidth underestimation under severe loss conditions. We thus perform experiments using the BBR~\cite{bbr} congestion control algorithm. BBR avoids underestimating the network bandwidth upon packet losses by looking at the receive rate and delay variation during the transfer. While not being explored in this paper, other congestion control algorithms~\cite{gcc,vegas} use other signals than packet losses to detect congestion.

\subsubsection{Experimental design}

We evaluate the bulk use-case by sending files of several sizes and first see how \solution{} compares with QUIC using its regular reliability mechanism. For this evaluation, we rely on an experimental design~\cite{experimental-design}. This approach consists in defining ranges of parameters instead of choosing precise values in order to mitigate the experimentation bias and explore network configurations showing the limits of the presented solution. We use the WSP~\cite{santiago2012construction} space-filling algorithm to cover the parameter space with 94 points. One experiment is run for each point in the parameter space.

\begin{figure*}
\begin{minipage}[t]{0.32\textwidth}
    \centering
    \includegraphics[trim=0.5cm 0cm 0.5cm 0.4cm,width=0.95\linewidth]{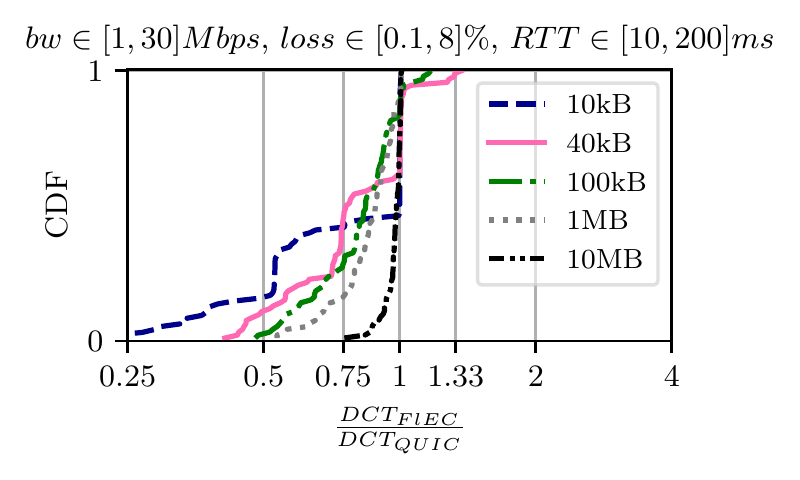}
    \caption{DCT ratio for bulk use-case using BBR. $\causalew{}()$ and $\causalth{}()$ ensure that Repair Symbols only protect the tail of the file.}
    \label{fig:results_bulk}
\end{minipage}
\hfill
\begin{minipage}[t]{0.32\textwidth}
    \centering
\includegraphics[trim=0.5cm 0cm 0.5cm 0.4cm,width=0.95\linewidth]{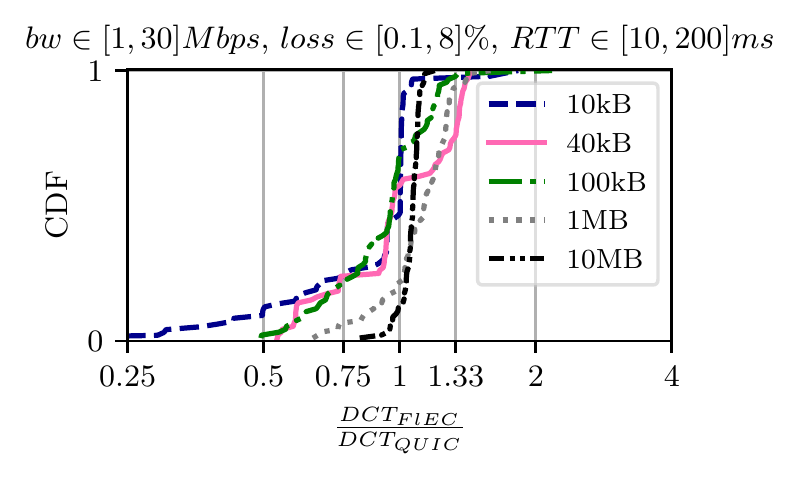}
    \caption{DCT ratio between \solution{} and AC-RLNC~\cite{cohen2020adaptive} for regular bulk use-case using the BBR congestion control.}
    \label{fig:results_bulk_flec_vs_ac_rlnc}
\end{minipage}
\hfill
\begin{minipage}[t]{0.32\textwidth}
    \centering
\includegraphics[trim=0.5cm 0cm 0.5cm 0.4cm,width=0.95\linewidth]{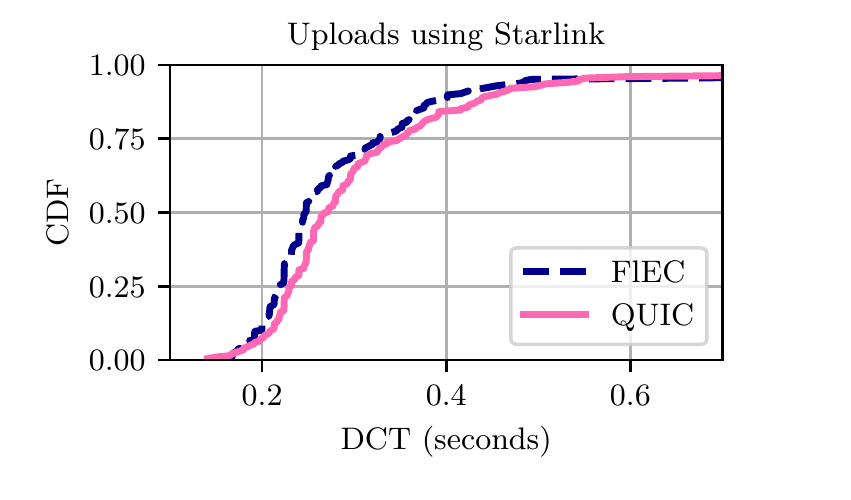}
    \caption{DCT comparing $\solution{}$ and the regular QUIC for downloads with at least one packet loss, performed on a real Starlink network access.}
    \label{fig:starlink_bulk_downloads_with_losses}
\end{minipage}
\end{figure*}

Figure~\ref{fig:results_bulk} shows the cumulative distribution function (CDF) of the Download Completion Time (DCT) ratio between \solution{} and \texttt{picoquic}~\cite{picoquic} used as our reference QUIC implementation. The experiments consist in the download files of size 10kB, 40kB, 100kB, 1MB and 10MB. For each file size, 95 experiments are run using experimental design. 40kB and 100kB are the average response sizes for Google Search on mobile and desktop devices~\cite{langley2017quic}. The parameter space is described on top of the Figure. The loss rate varies between 0.1\% and 8\% to cover both small loss rates and loss rates experienced under intense network conditions such as In-Flight Communications~\cite{rula2018mile}. The round-trip-time varies between 10ms and 200ms to experience both low delays and large delays such as those encountered in satellite communications. As shown in the Figure, $\causalth{}()$ and $\causalew{}()$ implement here a bulk-friendly reliability mechanism. By automatically protecting the tail of the downloaded file, we obtain similar results as previous works~\cite{de2019pluginizing,michel2019quic}. A few of the experiments with 40 and 100kB files provided poorer results compared to QUIC. With those file sizes, \solution{} uses one more stream frame to transmit the data, needing in some rare cases one additional round-trip to transmit this additional packet. While not shown graphically in this article, replacing BBR by Cubic~\cite{ha2008cubic} provides similar results. These experiments are provided in the artefacts that come with the article upon publication.

Figure~\ref{fig:results_bulk_flec_vs_ac_rlnc} compares $\solution{}$ with an implementation of AC-RLNC~\cite{cohen2020adaptive} following Table~\ref{tab:use_cases}. We observe that $\solution{}$ still outperforms AC-RLNC as sending repair symbols every RTT consumes too much bandwidth for the bulk use-case, while $\solution{}$ only sends repair symbols \emph{a priori} for the last flight of packets, relying on retransmissions for all the other packets as their retransmission arrives before the end of the download.

\subsubsection{Experimenting with a real network}
We now extend our study and analyze the benefits of \solution{} over a real network between a regular QUIC and \solution{} server on a Ubuntu 18.04 server located at UCLouvain and a client wired to a Starlink access point located in Louvain-la-Neuve (Belgium). We performed a total of 20150 uploads of 50kB from the client to the server. Among those 20150 uploads, 430 encountered at least one packet loss during the transfer. Figure~\ref{fig:starlink_bulk_downloads_with_losses} shows the CDF of the download completion time for these 430 uploads. The median download completion time for these uploads is 247ms for \solution{} and 272ms for regular QUIC. The average download completion time is 340ms for \solution{} and 393ms for QUIC. Unsurprisingly, \solution{} improves the download completion time for the transfers where the loss events occur during the RTT.

\subsubsection{CPU performance} While it has been demonstrated that PQUIC protocol plugins deteriorate noticeably the performance~\cite{de2019pluginizing}, we analyze the CPU impact of the \solution{} framework by transferring 1GB files on the loopback interface. Without \solution{}, we achieved a throughput of 650 Mbps. With \solution{} configured for the bulk use-case (i.e. sending Repair Symbols at the end of the transfer only), it dropped to 300 Mbps. This is inline with earlier observations on PQUIC performance. We believe that with a native implementation, the impact of the \solution{} framework would be barely noticeable.
We also analyzed the throughput sending one Repair Symbol every ten Source Symbols and obtained a throughput (i.e. not goodput) of 280 Mbps, meaning that the encoding and decoding of Repair Symbols implies only a small overhead compared to the framework in itself.

\section{Buffer-limited file transfers}\label{section:buffer_limited}
We here present and evaluate the reliability mechanism for buffer-limited file transfers. In this setup, the receive window (\textit{rwin}) is relatively small compared to the congestion window (\textit{cwin}) of the sender, making every loss event potentially blocking and increasing the download completion time. In addition to protect the download from tail losses, we protect every window of packets to avoid stalling due to lost packets blocking the stream flow-control window.

\subsection{Reliability mechanism}

For this use-case, $\causalth{}()$ returns $\hat{l}$ to ensure that $ad$ stays larger than $md$, according to the estimated loss rate.
$\causalew{}()$ behaves as shown in Algorithm~\ref{algo:buffer_ew}.
We spread the Repair Symbols along the sent Source Symbols in order to periodically allow the receiver to unblock its receive window by recovering the lost Source Symbols and deliver the stream data in-order to the application. More precisely, the $\causalew{}()$ operation sends one Repair Symbol every $\frac{1}{\hat{l}}$ Source Symbols. The algorithm needs three loss statistics. The first is the estimated uniform loss rate $\hat{l}$.
The two others are the $\hat{G_p}$ and $\hat{G_r}$ parameters of the Gilbert loss model. The Gilbert model~\cite{gemodel} is a two-states Markov model representing the channel, allowing representing network configurations where losses occur in bursts. These loss patterns cannot be easily recovered by a simple XOR error correcting code as shown in the original QUIC article~\cite{langley2017quic} but can be recovered by the random linear codes used by \solution{}. In the \textit{GOOD} state of the Gilbert model, packets are received while the packets are dropped in the \textit{BAD} state. $\hat{G_p}$ is the transition probability from the GOOD to the BAD state while $\hat{G_r}$ is the transition probability from the BAD to the GOOD state. In order to estimate the loss statistics $\hat{l}$, $\hat{G_p}$ and $\hat{G_r}$, we implement a \textit{loss monitor} that estimates the loss rate and Gilbert model parameters over a QUIC connection.

When the sender is blocked by the QUIC stream flow control, $\causalew{}()$ sends more Repair Symbols to recover from the remaining potentially lost Source Symbols. While spreading the Repair Symbols along the coding window helps to recover the lost Source Symbols more rapidly compared to a block approach where all the repair symbols are sent at the end of the window, this also potentially consumes more bandwidth. Indeed, the Repair Symbols do not protect the entire window. This means that with an equal number of losses, some specific loss patterns will lead to Repair Symbols protecting a portion of the window with no loss and portions of the window requiring more Repair Symbols to be recovered.

\begin{algorithm}\small
	\caption{\causalew{} for buffer-limited use-case}\label{algo:buffer_ew}
	\begin{algorithmic}[1]
		\Require $last$, the ID of the last symbol present in the coding window when $\causalew{}()$ was triggered the last time
		\Require $nTriggered$, the number of times \causalew{}() has already been triggered since no new symbol was added to the window.
		\Require $maxTrigger$, the maximum number of times we can trigger this threshold for the same window
		\Require $nRSInFlight$, the number of Repair Symbols currently in flight
		\Require $W$, the current coding window.
		\Require $FCBlocked()$, telling us if we are currently blocked by flow control.
		\Require $\hat{l}$, $\hat{G_p}$, $\hat{G_r}$, see Table~\ref{tab:symbols}.
		\If {$nRSInFlight \geq 2*\lceil|W|*\hat{l}\rceil$}
		    \State \Return $false$ \Comment{Wait for feedback before sending new RS}
	    \EndIf
		\State $nUnprotected \gets W.last - last$
		\State $n \gets min(\frac{1}{\hat{G_p}}, |W|)$
		\State $protect \gets nUnprotected = 0 \lor nUnprotected \geq n \lor FCBlocked()$
		\If{$protect \land nUnprotected \neq 0$}
	        \Comment{\textit{Start Repair Symbols sequence}}
	        \State $nTriggered \gets 1$
	        \State $last \gets W.last$
	        \State $maxTrigger \gets \lceil max(\hat{l}*nUnprotected, \frac{1}{\hat{G_r}}) \rceil$
	    \ElsIf{$protect$}
		    \If{$ FCBlocked() \lor nTriggered < maxTrigger$}
		
		      \State $nTriggered \gets nTriggered + 1$ \Comment{\textit{Continue sending symbols}}
		
		    \Else
		
		      \State $protect \gets false$ \Comment{\textit{Enough symbols have been sent}}
		    \EndIf
		\EndIf
		\State \Return $protect$
	\end{algorithmic}
\end{algorithm}

\subsection{Evaluation}
We now evaluate our generic mechanism under a buffer-limited file transfer use-case. We first study a specific network configuration that could benefit from \solution{}. We then evaluate its overall performance using  experimental design.

\subsubsection{\solution{} for SATCOM}
We choose the satellite communications (SATCOM) use-case where the delay can easily reach several hundreds of milliseconds~\cite{thomas2019quicsat,kuhn-quic-4-sat}. In those cases, end-hosts need a large receive buffer in order to reach the channel capacity. If they do not use a sufficiently large buffer, packet losses can have a significant impact on the throughput, preventing the sender to send new data as long as the data at the head of the receive buffer have not been correctly delivered to the application. The studied network configuration has a round-trip-time of 400 milliseconds and a bandwidth of 8~Mbps. Those are lower-bound values compared to current deployments~\cite{kuhn-quic-4-sat,thomas2019quicsat}. The bandwidth-delay product is thus 400kB. 
Higher BDP configurations are studied in the experimental design analysis of the next section.
We study the benefits brought by \solution{} with several receive window sizes.

\begin{figure*}
\begin{minipage}[t]{0.32\textwidth}
    \centering
\includegraphics[trim=0cm 0.7cm 0cm 1.7cm,width=\linewidth]{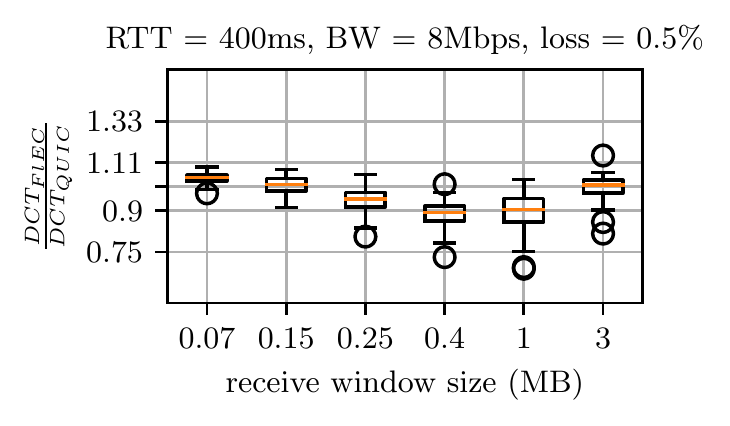}
    \caption{DCT ratio, 0.5\% losses.}
    \label{fig:buffer_bbr_05}
\end{minipage}
\hfill
\begin{minipage}[t]{0.32\textwidth}
    \centering
\includegraphics[trim=0cm 0.7cm 0cm 0cm,width=\linewidth]{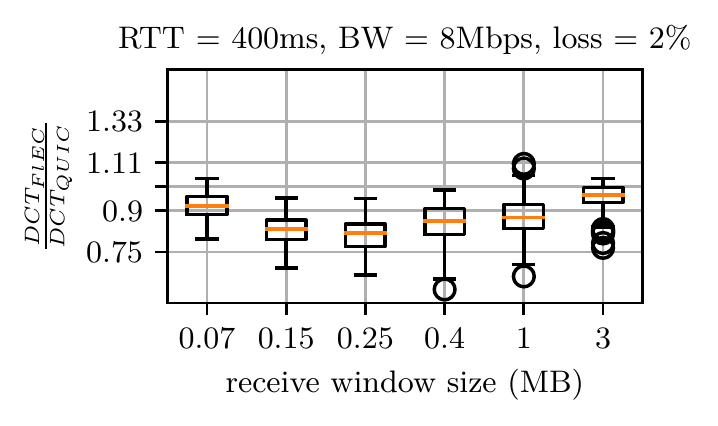}
    \caption{DCT ratio, 2\% losses}
    \label{fig:buffer_bbr_2}
\end{minipage}
\hfill
\begin{minipage}[t]{0.32\textwidth}

    \centering
\includegraphics[trim=0cm 0.7cm 0cm 1.7cm,width=\linewidth]{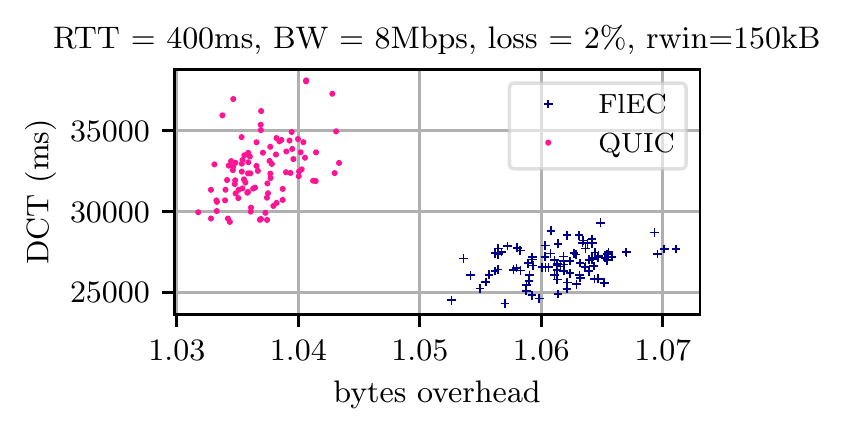} \caption{Time-bandwidth tradeoff, 2\% loss.}
    \label{fig:buffer_bytes_tradeoff_150kB}

\end{minipage}

\end{figure*}

\paragraph{Download completion time and throughput}
Figure~\ref{fig:buffer_bbr_05} shows the download completion time ratio between \solution{} and regular QUIC with a 5~MB file and 0.5\% of packet loss. Each box in the graph is computed from 95 runs with different seeds for the ns-3 rate error model. The bandwidth is set to 8~Mbps and the congestion control is BBR. For each transfer using \solution{}, we decrease the receive window by 5\% at the receiver in order to store the received repair symbols in the remaining space. With receive windows smaller than the BDP (ranging from 70~kB to 400~kB), the sender is flow-control-blocked once per RTT during a time proportional to the $\frac{rwin}{cwin}$ ratio. This implies that the download completion time with small receive windows is large even without any packet loss. When losses occur, the repair symbols sent \emph{a priori} help to unblock the receive window at the receiver-side and avoid blocking the data transfer for more than one RTT. For the 70~kB receive window, the 5\% reduction to store the repair symbols is significant compared to the benefit of FEC and has a negative impact on the goodput.
With the 400~kB receive window, the sender only blocks in the presence of losses during the round-trip. The earlier the loss occurs during the round-trip, the longer the sender will be blocked by the flow control for the next round-trip, since it needs to retransmit the data to unblock the receive window. Sending \emph{a priori} Repair Symbols for these configurations allows reducing or completely avoiding those blocking situations, at the price of a small reduction in goodput. The transmission of Repair Symbols in a sliding-window manner (i.e. interleaved with the Source Symbols) as described in Algorithm~\ref{algo:buffer_ew} helps to recover from losses earlier compared to sending all the Repair Symbols at once in a block fashion. The price to pay compared to a block pattern is an goodput reduction as some loss patterns might require more Repair Symbols to be recovered with this method. For the large receive windows, sending Repair Symbols \emph{a priori} does not unblock the window but still helps to recover from tail losses. With such a high RTT, the impact of a tail loss relative to the download completion time is still significant.

Figure~\ref{fig:buffer_bbr_2} shows the result of our experiments with a 2\% packet loss rate. It is thus more common that the sender becomes flow-control blocked. This makes the approach worth even for smaller receive window sizes such as 70kB as the sender will be slowed down a lot more often.


\paragraph{Delay-bandwidth tradeoff}
Figure~\ref{fig:buffer_bytes_tradeoff_150kB} illustrates the delay-bandwidth tradeoff operated when using \solution{} instead of regular QUIC. Each point on the figure concerns a single experiment and represents the download completion time and the bytes overhead of the solution. The bytes overhead is computed by dividing the total amount of bytes of UDP payload sent by the server by the size of the file transferred (5MB). For this graph, the experiments use a small receive window of 150kB and the loss rate is 2\%.
As the receive window is small, sending FEC unblocks the receive window upon losses and allows drastically lowering the download completion time. The price to pay is an additional bytes overhead compared to the regular QUIC solution. In this rwin-limited scenario, the available bandwidth is generally larger than what is used due to the rwin restriction.

\begin{figure*}
\begin{minipage}[t]{0.32\textwidth}
    \centering
 \includegraphics[trim=0cm 0.7cm 0cm 1.7cm,width=\linewidth]{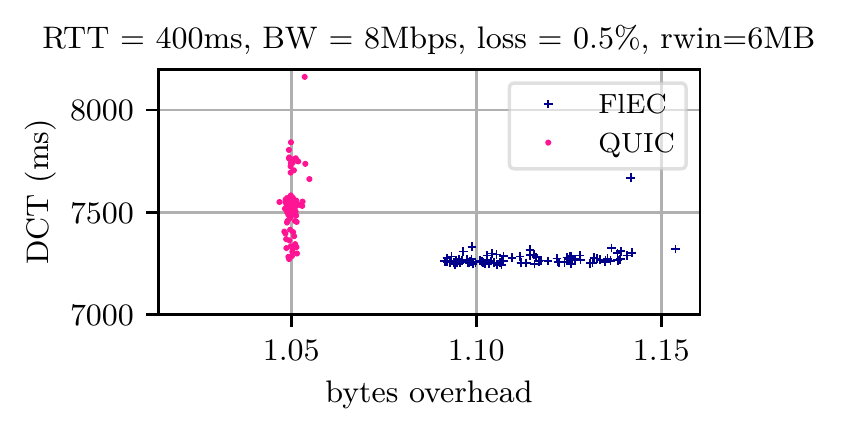}
     \caption{Time-bandwidth tradeoff with a 0.5\% loss link and a 6MB receive window.}
     \label{fig:buffer_bytes_tradeoff_6MB}
\end{minipage}
\hfill
\begin{minipage}[t]{0.32\textwidth}
    \centering
\includegraphics[trim=0cm 0.7cm 0cm 0cm,width=\linewidth]{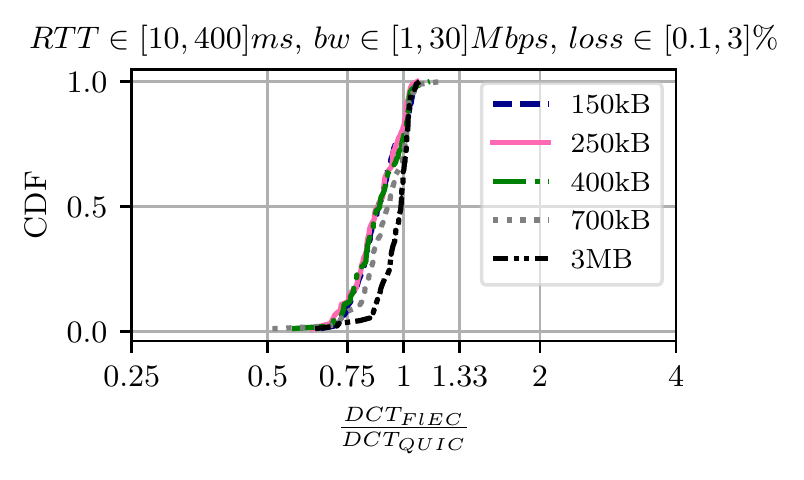}
    \caption{Experimental design analysis for several receive window configurations.}
    \label{fig:buffer_exp_design}
 \end{minipage}
 \hfill
\begin{minipage}[t]{0.32\textwidth}
    \centering
\includegraphics[trim=0cm 0.7cm 0cm 0cm,width=\linewidth]{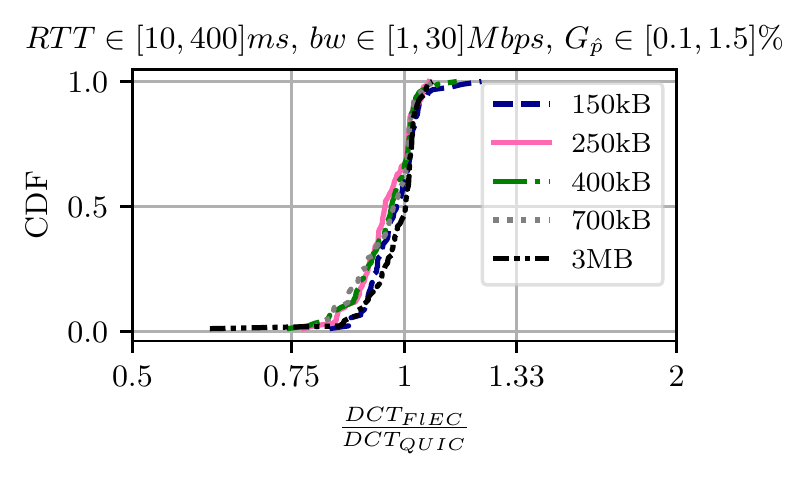}
    \caption{Experimental design analysis using Gilbert model with bursts of 1 to 5 packets.}
    \label{fig:buffer_exp_design_bursts}
 \end{minipage}

 \end{figure*}

Figure~\ref{fig:buffer_bytes_tradeoff_6MB} shows experiments results with the opposite scenario: the receive window is 6MB large, which is larger than both the file to transfer and the bandwidth-delay product of the link. This case is similar to the bulk use-case of section~\ref{section:bulk}. We can see that \solution{} leads to stable latency results at the expense of a larger bytes overhead. As the receive window is larger than the file to transfer, the sender will never be flow-control blocked during the download. In this case, \solution{} minimizes the latency essentially by recovering from tail losses.

\subsubsection{Experimental design analysis}

Figure~\ref{fig:buffer_exp_design} shows the aggregated results of simulations using experimental design. We show the CDF of the download completion time ratio between \solution{} and \texttt{picoquic}~\cite{picoquic}. Each CDF on the figure is built from 95 experiments with parameters selected from the ranges depicted on top of Figure~\ref{fig:buffer_exp_design}. Each CDF curve corresponds to downloads using the 
receive window size specified in the legend. The congestion control used is still BBR. 
We observe positive results using \solution{} for the majority (75\%) of the network configurations, especially for smaller receive window sizes (80\% positive results for windows smaller or equal to 400kB). Some configurations still expose negative results using \solution{}, even for smaller receive window sizes. These configurations are those whose bandwidth-delay product is small compared to the receive window. To verify this, we computed the average $\frac{BDP}{rwin}$ ratio on all the experiments for which \solution{} took more time to complete than \texttt{picoquic}, whose value is 0.48. For the experiments where the \solution{} download was faster, the average value of this ratio is 1.53.

Let us now assess the performance of our solution using a bursty loss model in order to see whether \solution{} stays robust even in presence of loss bursts. Figure~\ref{fig:buffer_exp_design_bursts} shows the results of an experimental design analysis with a Gilbert loss model with $G_{\hat{p}}$ ranging from 0.1\% to 1.5\% and $G_{\hat{r}}$ set to 33\% (i.e. an expected burst size of 3 packets) and a maximum burst size of 5 packets. Loss events thus occur less often compared to Figure~\ref{fig:buffer_exp_design}, leading to fewer blocking periods for QUIC during the experiments but with a higher probability of loosing several packets in a row. We can see that Algorithm~\ref{algo:buffer_ew} still offers benefits in the presence of bursty losses. Similarly to Figure~\ref{fig:buffer_exp_design}, \solution{} especially improves the results for experiments with a large $\frac{cwin}{rwin}$ ratio.

\section{Delay-constrained messaging}\label{section:messaging}

In this section, we present the implementation and evaluation of \solution{} tailored for delay-constrained messaging. The goal is to protect whole messages instead of naively interleaving Repair and Source Symbols. Using application knowledge, \solution{} protects as much frames as possible at once.

\subsection{Reliability mechanism}\label{section:messaging_mechanism}

We consider an application sending variable-sized messages, each having its own delivery deadline.
To convey these deadlines, we extend the transport API (Section~\ref{section:messaging_api}). Furthermore, we replace the QUIC stream scheduler to leverage application information (Section~\ref{section:messaging_scheduler}). This can be done easily since applications are bundled with their QUIC implementation and are able to easily extend it. 
We then discuss and evaluate a specific use-case in Section \ref{section:messaging_evaluation}.

\subsubsection{Application-specific API}\label{section:messaging_api}

We propose the following API enabling an application to send deadline-constrained messages.

\paragraph{\texttt{send\_fec\_protected\_msg(msg, deadline)}}
The application submits its deadline-constrained messages. The QUIC protocol already supports the stream abstraction as an elastic message service. However, the stream priority mechanism proposed by QUIC, while being dynamic, is not sufficient to support message deadlines.
The protocol operation attached to this function inserts the bytes submitted by the application in a new QUIC stream, closes the stream, and attaches the application-defined delivery deadline to it. 
The message must be delivered at the receiver within this amount of time to be considered useful.
If the network conditions prevent an on-time delivery of the message, the message may be cancelled, possibly before being sent and the underlying stream be reset.

\paragraph{\texttt{next\_message\_arrival(arrival\_time)}}

This API call allows the application to indicate when it plans to submit the next message. While this API function is not useful for all kinds of unreliable messaging applications, applications having a constant message sending rate such as video-conferencing might benefit from providing such information.

\subsubsection{Application-tailored stream scheduler}\label{section:messaging_scheduler}

The knowledge provided by the application to the transport layer is not only useful for the coded reliability mechanism. The information provided by the application-defined API calls is also valuable for the QUIC stream scheduler. Without this information, the QUIC scheduler schedules high priority streams first and has two different ways to handle the scheduling of streams with the exact same priority: $i)$ round-robin or $ii)$ FIFO.
We let the application define its own scheduler to schedule its streams more accurately. 
Algorithm~\ref{stream_scheduler} describes 
our QUIC stream scheduler for deadline-constrained messaging applications.

\begin{algorithm}\small
	\caption{Application-tailored scheduler for delay-constrained messaging.}\label{stream_scheduler}
	\begin{algorithmic}[1]
		\Require $ \mathcal S $, the set of available QUIC streams
		\Require $\hat{OWD}$, the estimated one-way delay of the connection
		\Require $now$, the timestamp representing the current time
		\Require $FCBlocked(stream)$, telling if the specified stream is flow control-blocked.
		\Require $closestDeadlineStream(\mathcal{S}, deadline)$, returning the non-expired stream with the closest delivery deadline to the specified deadline
		\State $scheduledStream \gets \varnothing$
		\State $currentDeadline \gets now + \hat{OWD}$ \Comment{Initialization}
		\While {$scheduledStream = \varnothing$}
			\State $candidate \gets closestDeadlineStream(\mathcal{S}, currentDeadline)$
			\If {$candidate = \varnothing$}
				\textbf{break}
			\EndIf
			\If {$\neg FCBlocked(candidate)$}
				\State $scheduledStream \gets candidate$
			\Else
			    \State $\mathcal{S} \gets \mathcal{S} \setminus \{candidate\}$
				\State $currentDeadline \gets candidate.deadline$
			\EndIf
		\EndWhile
		\If {$scheduledStream = \varnothing$}
		    \State \Return $defaultStreamScheduling(\mathcal{S})$ \Comment{Fallback}
		\EndIf
	\end{algorithmic}
\end{algorithm}

The $closestDeadlineStream()$ function searches among all the available streams attached to a deadline to find the stream having the closest expiration deadline while still having chances to arrive on-time given the current one-way delay. The scheduler chooses the non-flow-control blocked stream that is the closest to expire while still having a chance to be delivered on-time to the destination. Our implementation estimates the one-way delay as $\frac{RTT}{2}$. Other methods exist~\cite{frommgen2018multipathowd,huitema-quic-ts-06}. Recent versions of \texttt{picoquic} include a mechanism for estimating the one-way delay~\cite{picoquic} when the hosts clocks are synchronized. In the absence of clock synchronization, the estimated one-way delay can only be interpreted relatively, which helps to estimate the one-way delay variation but not for decision thresholds such as the one used in Algorithm~\ref{stream_scheduler}.

\subsubsection{\causalew{}() and \causalth{}() for delay-constrained messaging}
We now describe how our application redefines \solution{}. Our application is sensitive to the delivery delay of entire messages more than the in-order delivery delay of individual packets. We thus set the $\causalth{}()$ threshold to $-\hat{l}$ as it is useful to retransmit non-recovered lost packets that can still arrive on-time. $\causalew{}()$ is described in Algorithm~\ref{algo:ew_messages}. The algorithm triggers the sending of Repair Symbols to protect as many messages as possible according to the messages deadline and the next expected message timestamp if provided by the application. The rationale is the following. If the unprotected messages can wait for new messages to arrive before being protected, $\causalew{}()$ does not send Repair Symbols and waits for the arrival of new messages. Otherwise, Repair Symbols are sent to protect the entire window until it is considered fully protected. This idea of waiting for new messages before protecting comes from the fact that the messages can be small and sending Repair Symbols for each sent message can lead to a high overhead. By doing so, $\causalew{}()$ adapts the code rate according to the application needs.

\begin{algorithm}\small
	\caption{$\causalew{}()$ for delay-constrained messaging.}\label{algo:ew_messages}
	\begin{algorithmic}[1]
		\Require $ \mathcal S $, the set of available QUIC streams
		\Require $\hat{OWD}$, the estimated one-way delay of the connection
		\Require $now$, the timestamp representing the current time
		\Require $closestDL(\mathcal{S}, deadline)$, returning the message deadline that will expire the sooner from the specified deadline
		\Require $last$, the last protected message.
		\Require $nTriggered$, the number of times \causalew{} has already been triggered since no symbol was added to the window.
		\Require $maxTrigger$, the maximum number of times we can trigger this threshold for the same window
		\Require $nextMsg$ (is $+\infty$ if the message API is not plugged), the maximum amount of time to wait before a new message arrives.
		\Require $cwin$, $bif$, the congestion window and bytes in flight.
		\Require $\theta$ space to save in cwin for directly upcoming messages.
		\Require $\hat{l}$, $\hat{G_p}$, $\hat{G_r}$, see Table~\ref{tab:symbols}.
		\State $nextDL \gets closestDL(\mathcal{S}, max(now + \hat{OWD}, last.deadline))$
		\State $protect \gets (nextDL = \varnothing \lor now + \hat{OWD} + nextMsg + \epsilon \geq nextDL)$
		
		\State $nUnprotected \gets W.last - last$

		\If{$protect \land nUnprotected \neq 0$}
	        \Comment{\textit{Start Repair Symbols sequence}}
	        \State $nTriggered \gets 1$
	        \State $last \gets W.last$
	        \State $maxTrigger \gets \lceil max(\hat{l}*nUnprotected, \frac{1}{\hat{G_{r}}}) \rceil$
	    \ElsIf{$protect$}
		    \If{$nTriggered < maxTrigger$}
		
		      \State $nTriggered \gets nTriggered + 1$ \Comment{\textit{Continue sending}}
		
		    \Else
		      \State $protect \gets false$ \Comment{\textit{Enough symbols have been sent}}
		    \EndIf
		\EndIf

		\State \Return $appLimited() \land protect \land \frac{cwin}{bif} > \theta$
	\end{algorithmic}
\end{algorithm}

\subsection{Evaluation}
\label{section:messaging_evaluation}

\begin{figure*}

\begin{minipage}[t]{0.32\textwidth}
    \centering
    \includegraphics[trim=0cm 0.7cm 0cm 0cm,width=\linewidth]{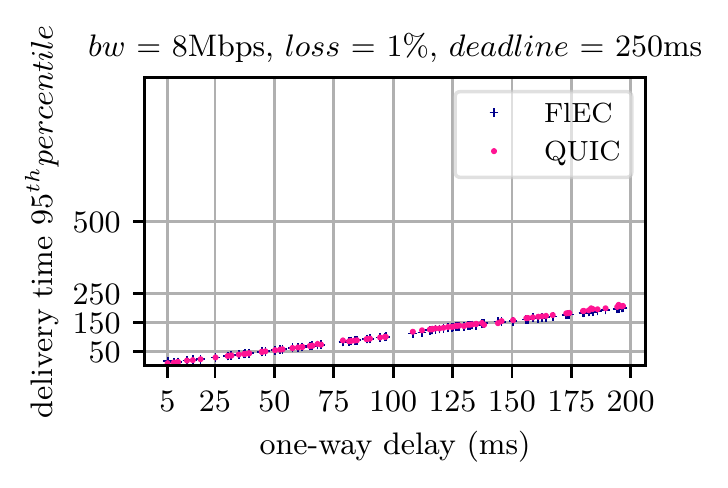}
    \caption{Message delivery time 95th percentile, comparing \solution{} with API and the regular QUIC.}
    \label{fig:p95_with_api_bbr}
\end{minipage}
\hfill
\begin{minipage}[t]{0.32\textwidth}
    \centering
\includegraphics[trim=0cm 0.7cm 0cm 0cm,width=\linewidth]{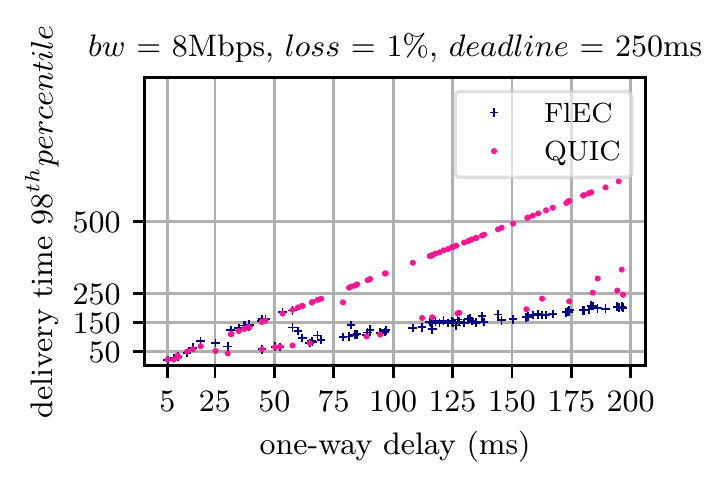}
    \caption{Message delivery time 98th percentile, comparing \solution{} with API and the regular QUIC.}
    \label{fig:p98_with_api_bbr}
\end{minipage}
\hfill
\begin{minipage}[t]{0.32\textwidth}
    \centering
\includegraphics[trim=0cm 0.7cm 0cm 0cm,width=\linewidth]{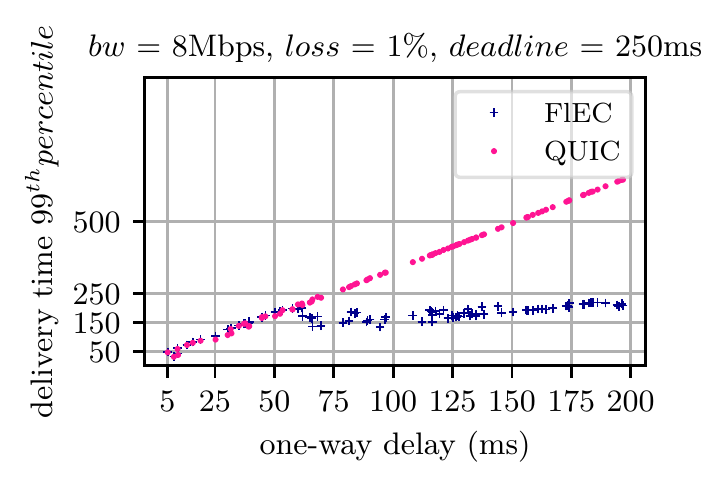}
    \caption{Message delivery time 99th percentile, comparing \solution{} with API and the regular QUIC.}
    \label{fig:p99with_api_bbr}
\end{minipage}

\end{figure*}

\begin{figure*}
\begin{minipage}[t]{0.32\textwidth}
    \centering
\includegraphics[trim=0cm 0.7cm 0cm 0cm,width=\linewidth]{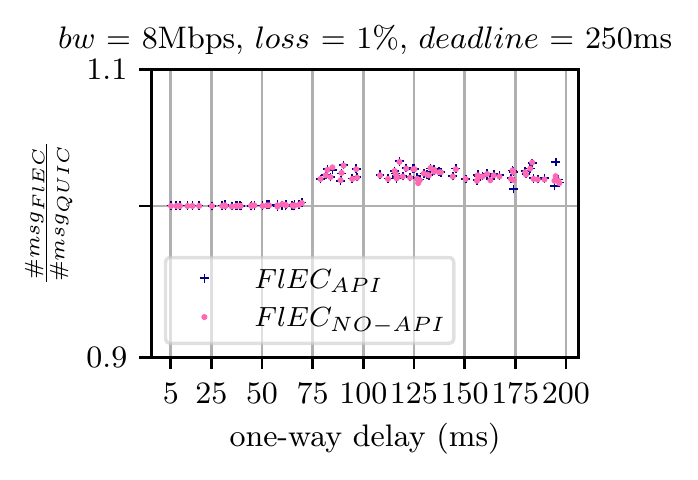}
    \caption{Messages received on-time comparing QUIC and \solution{} with and without API, using BBR.}
    \label{fig:messages_ratio_with_and_without_api}
\end{minipage}
\hfill
 \hfill
\begin{minipage}[t]{0.32\textwidth}
    \centering
\includegraphics[trim=0cm 0.7cm 0cm 0cm,width=\linewidth]{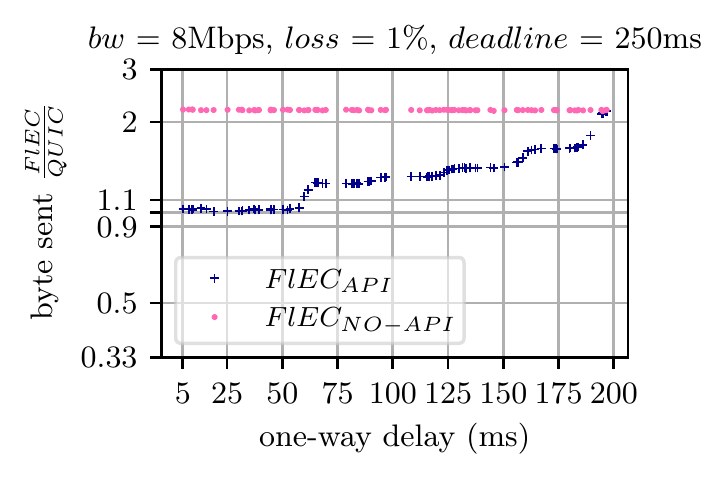}
    \caption{Bytes sent by the server, comparing QUIC and \solution{} with and without API, using BBR.}
    \label{fig:bytes_sent_with_and_without_api}
\end{minipage}
 \hfill
\begin{minipage}[t]{0.32\textwidth}
    \centering
\includegraphics[trim=0cm 0.7cm 0cm 0cm,width=\linewidth]{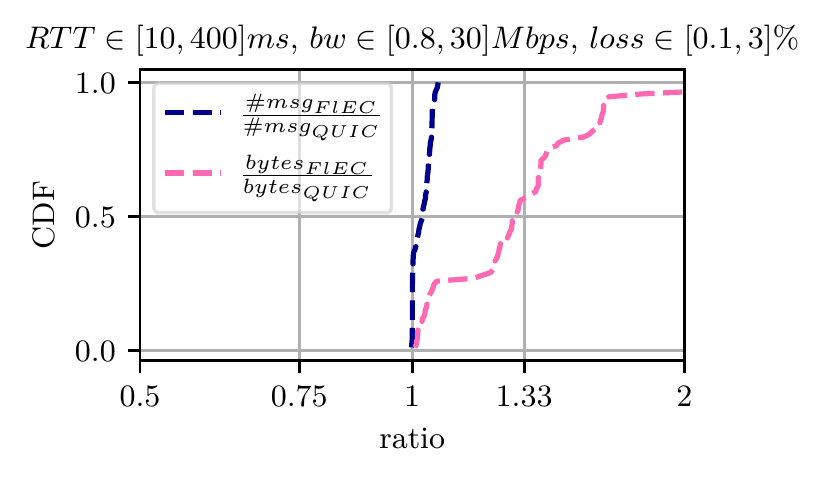}
    \caption{Experimental design analysis for the delay-constrained messaging use-case using BBR.}
    \label{fig:message_experimental_design}
\end{minipage}
\end{figure*}

We evaluate \solution{} under the messaging use-case using an application sending video frames as messages. 
We set the deadline to 250 milliseconds, meaning that each frame must be delivered within this time. We use 86 seconds of the video recording from the Tixeo video-conference application~\cite{tixeo}. The framerate and bitrate are adjusted by the application. This video recording starts at 15 frames per second during the first 6 seconds and runs at 30 images per second afterwards. For frame, we record its delivery delay between when the application sends it and when it is delivered at the receiver.
We send each video frame in a different QUIC stream to avoid head-of-line blocking across frames upon packet losses. 
The regular QUIC solution uses the default round-robin scheduler provided by PQUIC.
In the first set of experiments, we set the bandwidth to 8~Mbps and observe the performance of \solution{} in the presence of losses. For each experiment, the delay is sampled in the $[5, 200]ms$ range. We then perform an experimental design analysis over a wider parameters space.

Figure~\ref{fig:p95_with_api_bbr} and Figure~\ref{fig:p98_with_api_bbr} show the 95$^{th}$ and 98$^{th}$ percentiles of the message delivery times for each experiment.
We can see that while 95\% of the video frames are delivered successfully in every experiment, regular QUIC struggles to deliver 98\% of the submitted frames on time (i.e. before 250 milliseconds) with a one-way delay above 75 milliseconds.
Indeed, with a one-way delay above 75 milliseconds, the lost frames are retransmitted after more than 150 milliseconds and take more than 75 milliseconds to reach the receiver. Note that QUIC's loss detection mechanism takes a bit more than one RTT to consider a packet as lost to avoid spurious retransmissions due to reordering~\cite{ietf-quic-recovery}. These retransmitted frames thus arrive a few milliseconds before the deadline in the best case.
As we can see on Figure~\ref{fig:p98_with_api_bbr}, only a few experiments without \solution{} have more than 98\% of the frames arriving on-time while \solution{} can cope with one-way delays up to 200 milliseconds. Figure~\ref{fig:p99with_api_bbr} shows that no experiment with regular QUIC succeeded to deliver 99\% of the video frames on time with a one-way delay above 75ms, while \solution{} succeeded in every experiment.

Note that the $\causalew{}()$ algorithm plugged in this use-case tries to protect as many messages as possible with the same number of Repair Symbols by delaying the sending of Repair Symbols when new messages are expected soon.
This lazy Repair Symbol scheduling explains the plateau present around the 250ms delivery time in Figure~\ref{fig:p99with_api_bbr} and why the frame delivery time is larger than the one-way delay. In order to send as few Repair Symbols as possible, \solution{} delays the sending of Repair Symbols to the last possible moment while ensuring that lost data can be recovered before the deadline.

Figure~\ref{fig:messages_ratio_with_and_without_api} shows the ratio between the number of messages received on-time by \solution{} and by the regular QUIC implementation. In order to isolate the effects of the \solution{} API, the Figure also shows \solution{} results without leveraging the application knowledge brought by the API functions ($FlEC_{NO-API}$ on the Figure). It thus uses \texttt{picoquic}'s default stream scheduler and sends repair symbols for each newly sent message. As it is only a simplified version of Algorithm~\ref{algo:ew_messages}, we do not show the $\causalew{}()$ algorithm of this second solution. As we can see, nearly no experiment ended with fewer messages received on-time using the API-enabled \solution{} compared to QUIC. A similar gain compared to the regular QUIC is present for both \solution{} versions. However, the interest of the \solution{} API resides in the redundancy it needs to obtain those results.

We now analyze the redundancy overhead of our solution. Figure~\ref{fig:bytes_sent_with_and_without_api} shows the ratio of bytes sent by the server between regular QUIC and \solution{} with and without the API defined in Section~\ref{section:messaging_api}. The results of \solution{} without the API show that protecting every message blindly is very costly in terms of bandwidth. Indeed, for this video-conferencing transfer, many video frames sent by the application are smaller than the size of a full QUIC packet. The QUIC \texttt{REPAIR} frames sent by \solution{} contain additional metadata. In this case where the application traffic is thin, protecting every message may double the volume of sent data as shown on Figure~\ref{fig:bytes_sent_with_and_without_api}. Using \solution{} with the message-based API can save a lot of bandwidth by using application-aware stream and redundancy schedulers.
Note the portion of the graph between 5ms and 70ms one-way delays. For those configurations, no Repair Symbol is sent by \solution{}. Indeed, the messages are acknowledged by the peer before  $\causalew{}()$ triggers the sending of repair symbols. \solution{} thus naturally uses SR-ARQ when redundancy is not needed to meet the messages deadlines.

\subsubsection*{Experimental design analysis}

Figure~\ref{fig:message_experimental_design} shows the results of an experimental design analysis using the parameters depicted on the top of the Figure. The Figure shows CDFs for the amount of bytes sent by the server and the number of messages received within the deadline. In a few cases, the \solution{} solution using the application-tailored API sends a similar amount of bytes to regular QUIC. This is due to the fact that for some configurations, the delay was sufficiently low to send no or a few repair symbols. We can also see that none of the experiments revealed a lower amount of on-time received messages compared to regular QUIC, showing the robustness of \solution{} under various network conditions.

\subsubsection*{Improvements}

Other information from the application could have been taken in addition to the messages deadline. For example, information concerning the video frames type could have an impact on the stream scheduling: H264 I-frames are more important than P as the latter depend on the first to be decoded. The stream scheduling can even be further improved by looking at the dependence between each frames in a group of H264 frames. Given the flexibility of \solution{}, the messaging API can be easily extended for the application to transfer this kind of knowledge to the transport stack. 
\section{Conclusion}

In this paper, we redefine the QUIC reliability mechanism and enable its per-use-case customization. Flexible Erasure Correction (\solution{}) allows efficiently combining retransmissions and Forward Erasure Correction.
Applications can either use a standard Selective-Repeat ARQ mechanism or tailor a Forward Erasure Correction mechanism that fits their own traffic pattern and sensitivity to delays. Our \solution{} implementation leverages the PQUIC protocol plugins to enable the application to insert its own algorithm to select the level of redundancy and the stream scheduling decisions. We customize \solution{} for three different use-cases. We evaluate and demonstrate that \solution{} can be configured with small to no effort by applications to significantly enhance the quality of experience compared to the existing QUIC loss recovery mechanisms. \solution{} is currently a single-path implementation. In the future, we plan to study how \solution{} can be used together with several network interfaces to improve the transfer for the considered use-case and go further than reducing head-of-line blocking using a tailored redundancy and path scheduler.


\section*{Artefacts}
Simulation scripts and the code of \solution{} are publicly available from \url{https://github.com/francoismichel/flec}.

\ifCLASSOPTIONcaptionsoff
  \newpage
\fi



\bibliographystyle{IEEEtran}

\bibliography{reference,rfc}
%

%

\begin{IEEEbiography}{Michael Shell}
Biography text here.
\end{IEEEbiography}

\begin{IEEEbiographynophoto}{John Doe}
Biography text here.
\end{IEEEbiographynophoto}


\begin{IEEEbiographynophoto}{Jane Doe}
Biography text here.
\end{IEEEbiographynophoto}




\end{document}